# Measuring and modeling polymer gradients argues that spindle microtubules regulate their own nucleation

Bryan Kaye[a,b], Olivia Stiehl[a,b], Peter J. Foster[a,b], Michael J. Shelley[d,e], Daniel J. Needleman[a,b,c], Sebastian Fürthauer[d]

[a]John A. Paulson School of Engineering and Applied Science, [b]FAS Center for Systems Biology, [c]Department of Molecular and Cellular Biology, Harvard University, Cambridge, MA 02138, [d]Center for Computational Biology, Flatiron Institute, New York, NY 10010, [e]Courant Institute, NYU, New York, NY 10012

## 0 Abstract

Spindles are self-organized microtubule-based structures that segregate chromosomes during cell division. The mass of the spindle is controlled by the balance between microtubule turnover and nucleation. The mechanisms that control the spatial regulation of microtubule nucleation remain poorly understood. Previous work has found that microtubule nucleators bind to microtubules in the spindle, but it is unclear if this binding regulates the activity of those nucleators. Here we use a combination of experiments and mathematical modeling to investigate this issue. We measure the concentration of tubulin and microtubules in and around the spindle. We found a very sharp decay in microtubules at the spindle interface, which is inconsistent with the activity of microtubule nucleators being independent of their association with microtubules and consistent with a model in which microtubule nucleators are only active when bound to a microtubule. This strongly argues that the activity of microtubule nucleators is greatly enhanced when bound to microtubules. Thus, microtubule nucleators are both localized and activated by the microtubules they generate.

## 1 Introduction

The spindle is a self-organized cellular structure which separates chromosomes during cell division. In meiotic Xenopus egg extract spindles, the spatial regulation of microtubule nucleation is crucial for establishing spindle architecture (Brugués et al. 2012; Wieczorek et al. 2015; Oh, Yu, and Needleman 2016; Decker, Oriola, Dalton, Brugues, bioRxiv, 174078). The Ran pathway, the same pathway used in interphase for nuclear import, promotes microtubule nucleation near chromosomes by activating spindle assembly factors (Ohba et al. 1999; Zhang, Hughes, and Clarke 1999; Kalab, Pu, and Dasso 1999). Many of these spindle assembly factors bind to microtubules (Petry et al. 2013; Kamasaki et al. 2013; Hsia et al. 2014; Ho et al. 2011; Wieczorek et al. 2015; Roll-Mecak and Vale 2006; Oh, Yu, and Needleman 2016). Outside of the spindle in Xenopus egg extracts, where individual microtubules can be visualized, it has been found that nucleators, and the microtubules which grow from them, localize to other microtubules in a "branching" pattern (Petry et al. 2013). While this demonstrates that nucleators can bind to preexisting microtubules, it remains unclear whether this binding stimulates the activity of nucleators.

At least two scenarios are possible. In one, taken in its extreme form, nucleators can only create new microtubules when bound to preexisting microtubules, while in the other, nucleator activity is unaffected by being bound to a preexisting microtubule (see Figure 1, top). Distinguishing between these two scenarios is difficult in the bulk of the spindle, since microtubules are present at near constant density, and thus it is hard to disentangle the binding-dependence of nucleator activity. However, we hypothesized that near the edge of the spindle, where the concentration of microtubules drops, the difference in nucleator and microtubule diffusion rates would result in a zone in which concentrations of microtubules and nucleators would vary substantially, which would allow us to discriminate the two scenarios.

To do this, we first quantified the decay of microtubule density at the spindle interface. Due to potential gradients in monomer concentration at the spindle interface, we could not rely on fluorescence microscopy alone to determine the polymer concentration. Instead, we used a FLIM-FRET based technique introduced by (Kaye et al. 2017), which allows separate measurements of the gradients of microtubules and tubulin monomers in and around spindles. Using this technique, we found that the microtubule concentration decreased sharply around the spindle.

We next developed a mathematical model for the shape of the Xenopus metaphase spindle interface. Since microtubules move relatively slowly ($\simeq 2\mu$m/min) and turn over rapidly with a lifetime of $\simeq$ 20s (Sawin 1991; Needleman et al. 2010), we ignored transport and only considered the reaction diffusion dynamics of microtubules and nucleators. If nucleators must be bound to preexisting microtubules to nucleate, then our model predicts a sharp microtubule gradient at the spindle boundary set by the distance microtubules diffuse before they depolymerize (Figure 1, left). If nucleator activity is unaffected by binding to preexisting microtubules, the model predicts the width of the spindle's interface to be broader because of the diffusion of nucleators (Figure 1, right). By comparing the expectations from the model to experimental results, we argue that nucleators are activated by binding to preexisting microtubules. This finding demonstrates that the spindle, which turns over rapidly and has no enclosing membrane, maintains a sharp interface by a feedback mechanism between the nucleator and its nucleation product.

The structure of our paper is as follows: In section 2 we give an overview of the technique used to measure microtubule mass and show that the microtubule gradient decays quickly away from the spindle. We also show that the shape of the interface is very similar from spindle to spindle and does not depend on the position along the spindle's circumference. In section 3 we outline the key ideas underlying our study and formulate a description of the reaction-diffusion dynamics that shape the spindle interface. We then combine theory and experiment to demonstrate that the shape of the spindle interface is consistent with nucleators being activated by binding to preexisting microtubules, but inconsistent with nucleators indiscriminately nucleating microtubules. Finally, in section 4 we consider the implications of these findings for our understanding of the spindle and other organelles.

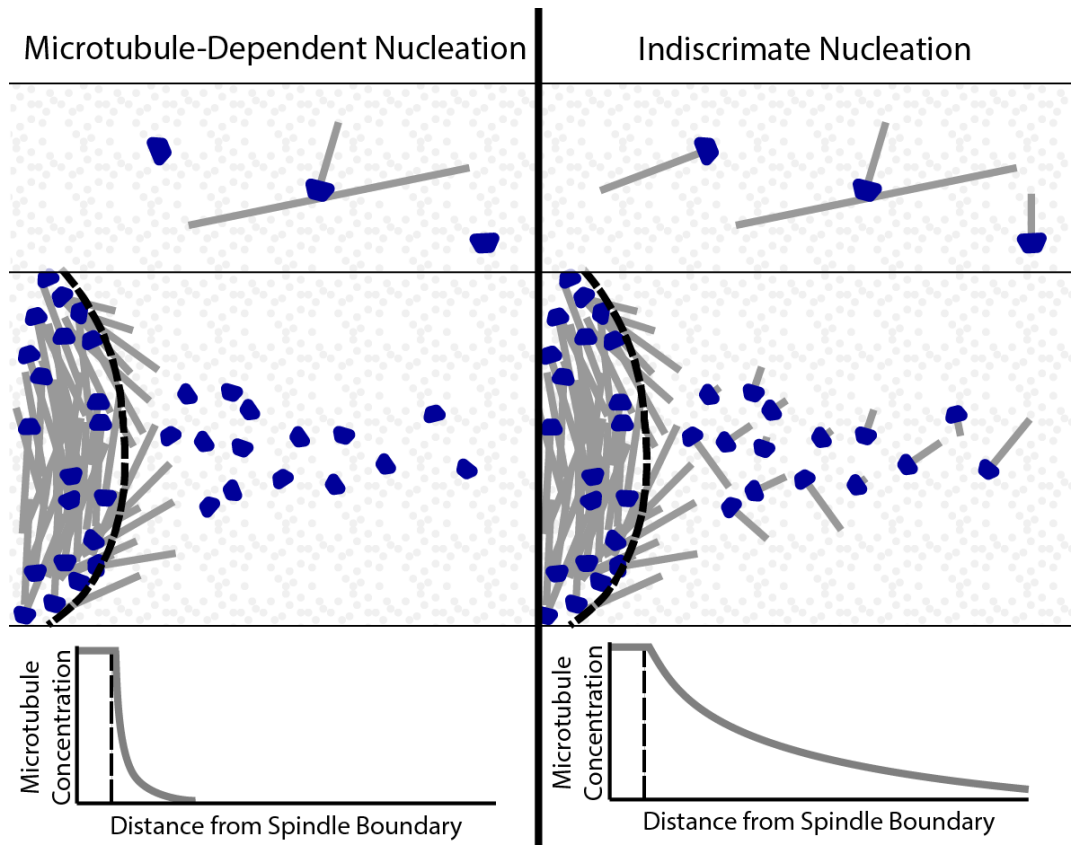

*Figure 1: (Top) Schematic representation of two extreme scenarios of nucleation. In the microtubule-dependent nucleation scenario, nucleators can only nucleate when bound to preexisting microtubules. In the indiscriminate nucleation scenario, nucleator activity is not affected by other microtubules. (Middle) Schematic representation of the spindle, which has a soft boundary and nucleators diffusing away from the spindle boundary. (Bottom) In the microtubule dependent nucleation scenario, the gradient in the microtubule concentration away from the spindle boundary is the interface width. In the indiscriminate nucleation scenario, nucleators continue to create microtubules as they diffuse away from the spindle. This predicts a much wider gradient in the microtubule concentration.*

## 2 Measuring the spindle interface

We sought to test models of microtubule nucleation by measuring how the concentration of microtubules decays at the spindle interface. We now describe how we performed these measurements and present the results.

### 2.1 FLIM-FRET

To measure the microtubule concentration at the spindle interface, we need a technique that can measure polymer concentration in and around spindles. If fluorescence microscopy were able to visualize each individual filament, then it could be used to measure the concentration of polymer. However, it is not possible to directly visualize individual microtubules in and around spindles because of their high density and the large background signal from soluble tubulin. A possible alternative approach for using fluorescence microscopy to measure microtubule concentration is to note that the fluorescence signal in each pixel is proportional to the total amount of labeled tubulin, which is the sum of soluble tubulin and tubulin in microtubules, in the voxel corresponding to that pixel. Thus, if the signal from soluble tubulin at every pixel were known, it could be subtracted from the fluorescence signal, revealing the concentration of microtubules. Using this approach requires an estimate of the soluble concentration of tubulin, which might be obtained by taking the fluorescence intensity far away from the spindle. However, that procedure is only valid if the concentration of microtubules far from the spindle is negligible and if the concentration of soluble tubulin is spatially uniform. It is not clear that either assumption is valid. The concentration of soluble tubulin could be depleted in spindles, because of extensive microtubule polymerization, or on the contrary be enriched, due to complex interactions between tubulin monomers and microtubules. Thus, standard fluorescence microscopy cannot be used to measure the spatial variation in microtubule concentration at the spindle interface. To overcome these challenges, we use both fluorescence microscopy and spectroscopy to determine the concentration of microtubules in each pixel without making any assumptions on the spatial distribution of soluble tubulin or concentration of microtubules far from the spindle.

We use the method introduced by (Kaye et al. 2017) to measure microtubule concentration. We provide a brief summary here: The measurement system relies on Fluorescence Lifetime Imaging Microscopy (FLIM) in the presence of two subpopulations of fluorophores which can engage in FRET (Förster Resonance Energy Transfer). Populations of tubulin, labeled with either donor or acceptor fluorophores, are added to Xenopus egg extract. FRET can occur when a donor and acceptor are within ≈ 5 nm of each other. This is exceedingly rare in solution. In contrast, when the labeled tubulin is incorporated into microtubules, the microtubule lattice localizes donors and acceptors together and FRET becomes more likely.

To measure the subpopulation of donors engaged in FRET and the subpopulation of donors not engaged in FRET, we use time-domain FLIM. To make FLIM measurements, the donor fluorophores are put into an excited state by a laser pulse and relax back to their ground state either by emitting a photon or by dissipating the energy as heat. The amount of time spent in the excited state is called the fluorescence lifetime. If an acceptor is nearby, an additional pathway of relaxation, FRET, is available, shortening the amount of time the donors spend in an excited state (Figure 1A) and thus shortening the fluorescence lifetime. If there is a subpopulation of donors engaged in FRET and a subpopulation not engaged in FRET, the measured photon emission will be the weighted sum of the photon emission from each subpopulation (Figure 1B), where the weights are the number of donors in each subpopulation. Thus, these weights can be deduced by analyzing the photon emission from a sample with both subpopulations, allowing the fraction of donors engaged in FRET to be measured, from which the fraction of donors in microtubules can be calculated.

Since fluorescence intensity provides a measure of the total amount of tubulin, we combine intensity measurements with simultaneous measurements of FRET fraction, which gives the fraction of tubulin in microtubules, to calculate the total amount of tubulin in microtubules. This is derived in (Kaye et al. 2017), where and the amount of donors in polymer (i.e. in microtubules), $N_{pol}(x)$, was shown to be related to FRET and intensity by:

$$N_{pol}(x) = \frac{I(x)F(x)}{bP_f(1 + (a-1)F(x))} \quad (1)$$

where $F(x)$ is the fraction of donors engaged in FRET at location $x$, $I(x)$ is the intensity at location x, $b$ is the average number of photons per donor not engaged in FRET, $P_f$ is the probability a donor can engage in FRET and $a$ is the relative brightness of donors engaged in FRET to donors not engaged in FRET. Similarly, the monomer concentration (i.e. tubulin not in microtubules), $N_{mon}(x)$, can be written as

$$N_{mon}(x) = \frac{I(x)\left(P_f - F(x)\right)}{bP_f(1 + (a-1)F(x))} \quad (2)$$

Thus, measuring both FRET (via FLIM) and fluorescence intensity in a spatially resolved manner provides a powerful tool to characterize the spindle interface.

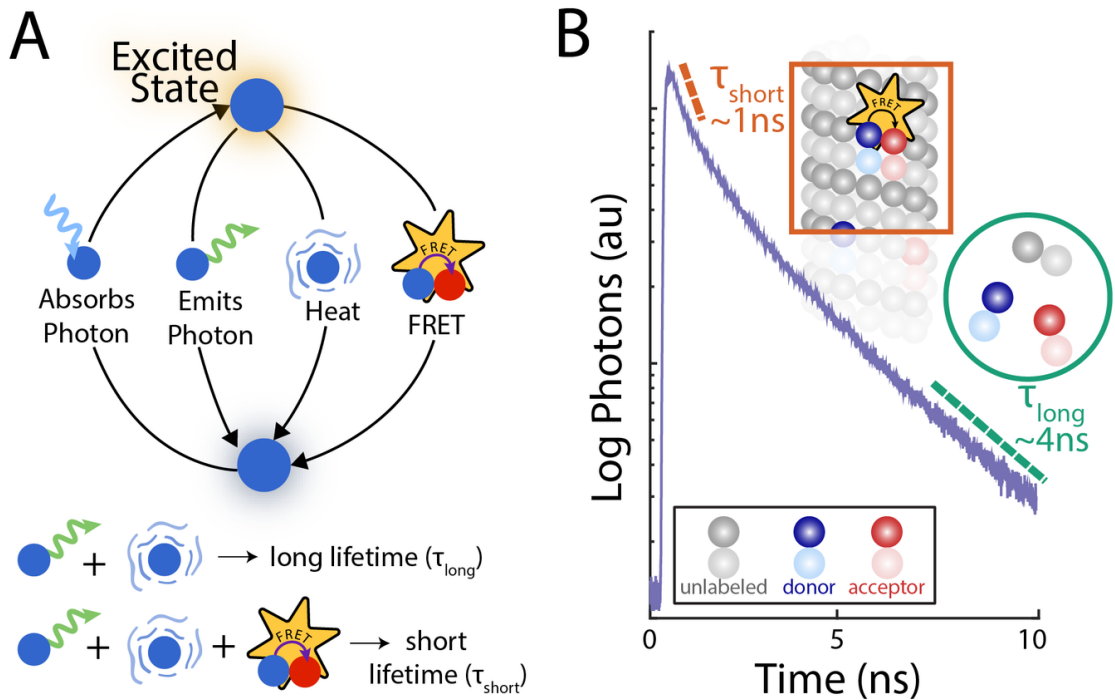

*Figure 2: FLIM measurements of FRET reveal the fraction of tubulin in polymer. (A) Diagram of the excitation and relaxation pathways of donor fluorophores (blue). If a donor fluorophore absorbs an incoming photon, it is raised into an excited state. The fluorophore relaxes back to the ground state by either emitting a photon or releasing heat. When an acceptor is nearby, FRET can occur, providing an additional pathway for the donor fluorophore to relax. Thus, the average time the fluorophore spends in an excited state, referred to as the lifetime, is shorter when the fluorophore is engaged in FRET. (B) Photon emission measurements, provided by FLIM, reveal the fraction of tubulin in polymer. Soluble donor-fluorophore-*

*labeled tubulin is very unlikely to be close enough to an acceptor to be engaged in FRET, and thus produces a long lifetime photon emission. Donor-labeled tubulin in microtubules can be in close proximity to acceptor-labeled tubulin, which produces a short-lifetime photon emission. By analyzing the photon emission curves, we can measure the ratio of tubulin in microtubules to soluble tubulin. Figure reproduced from (Kaye et al. 2017)*

## 2.2 Shape of the spindle interface

We next measured the microtubule gradient from the spindle boundary. We first added donor- and acceptor-labeled tubulin to Xenopus egg extract containing metaphase spindles (Experimental Methods). We then imaged spindles and defined the spindle boundary by thresholding the intensity image, and segmented the image into pixels that were equidistant from the spindle boundary (Experimental Methods, Figure 3A). Pixels within the spindle boundary are defined to have a negative distance, while pixels outside the boundary are defined to have a positive distance. The pixels were then binned according to distance and the photons corresponding to these pixels were analyzed to find the FRET fraction and intensity (Figure 3B). The calculated FRET fraction was largest in the spindle and decayed outside the spindle to $0.044 \pm 0.004$, where we define outside the spindle to mean further than 5 microns from the interface. The finite FRET fraction measured outside the spindle might be caused by the presence of microtubules in solution far removed from the spindle, or it could be an artifact caused by errors in the FLIM measurement (perhaps due to mischaracterization of the instrument response function, auto-fluorescence of the extract, or slight deviations of the donor lifetime distribution from a single exponential).To estimate the extent to which the calculated FRET fraction is not due to actual FRET, we measured the intensity and FRET fraction from spindles that only contained donor-labeled tubulin (and thus could not produce FRET because of the absence of acceptor-labeled tubulin). Averaging FRET and intensity measurements from 3 spindles revealed a similar donor intensity profile to spindles with acceptor-labeled tubulin incorporated. In contrast, the calculated FRET fraction in the absence of acceptor-labeled tubulin was spatially uniform, with an average value of $0.046 \pm 0.003$.

As the donor-labeled tubulin cannot engage in FRET in the absence of acceptor-labeled tubulin, the spatially uniform calculated FRET fraction must be due to an artefactual offset. An artefactual offset in measured FRET fraction was previously seen with free dye in buffer (Kaye et al. 2017). To correct for this artifact, we averaged the FRET fraction over distance in samples that lacked acceptor-labeled tubulin, and used this value to correct the artefactual FRET fraction offset in samples with acceptor-labeled tubulin. We calculate a corrected polymer and monomer concentration using Eqs 1 and 2 and obtain:

$$N_{pol}(x) = \frac{I(x)(F(x) - F_0)}{bP_f\big(1 + (a-1)(F(x) - F_0)\big)} \quad (3)$$

$$N_{mon}(x) = \frac{I(x)\big(P_f - (F(x) - F_0)\big)}{bP_f\big(1 + (a-1)(F(x) - F_0)\big)} \quad (4)$$

where $F_0$ is the artefactual offset. $a$, the relative brightness of donors engaged in FRET to donors not engaged in FRET, was estimated from the ratio of the fluorescence lifetimes as previously described (Kaye et al. 2017). To determine $b$, the brightness of donors not engaged in FRET per μM tubulin, we measured the average intensity far from the spindle (>10μm) and assume the tubulin concentration in this region to be 18 μM ((Parsons and Salmon 1997), Experimental Methods). $P_f$ is measured by grouping pixels by intensity, and then fitting the relationship between FRET and intensity as previously described (Kaye et al. 2017). We then calculated the polymer concentration from the FRET and intensity measurements shown in Figure 3B with and without the offset $F_0$ using Eqs. 1 and 2, respectively (Figure 3D). In both cases, we see that the concentration of tubulin in microtubules quickly decays from a maximum value in the spindle, to a low level outside the spindle. If the artefactual FRET offset is not corrected for, the microtubule concentration outside the spindle would be estimated to 7.96 $\mu M \pm 0.76 \mu M$ which corresponds to approximately 10% of its maximum value (Figure 3D, grey datapoints). After correcting for the offset, the microtubule concentration we determine outside the spindle is $-0.21 \mu M \pm 0.72 \mu M$ (Figure 3D, black datapoints), indicating a negligible amount of the spontaneous nucleation of microtubules outside the spindle. These measurements reveal that the microtubule concentration outside the spindle is less than 1.5μM (Methods).

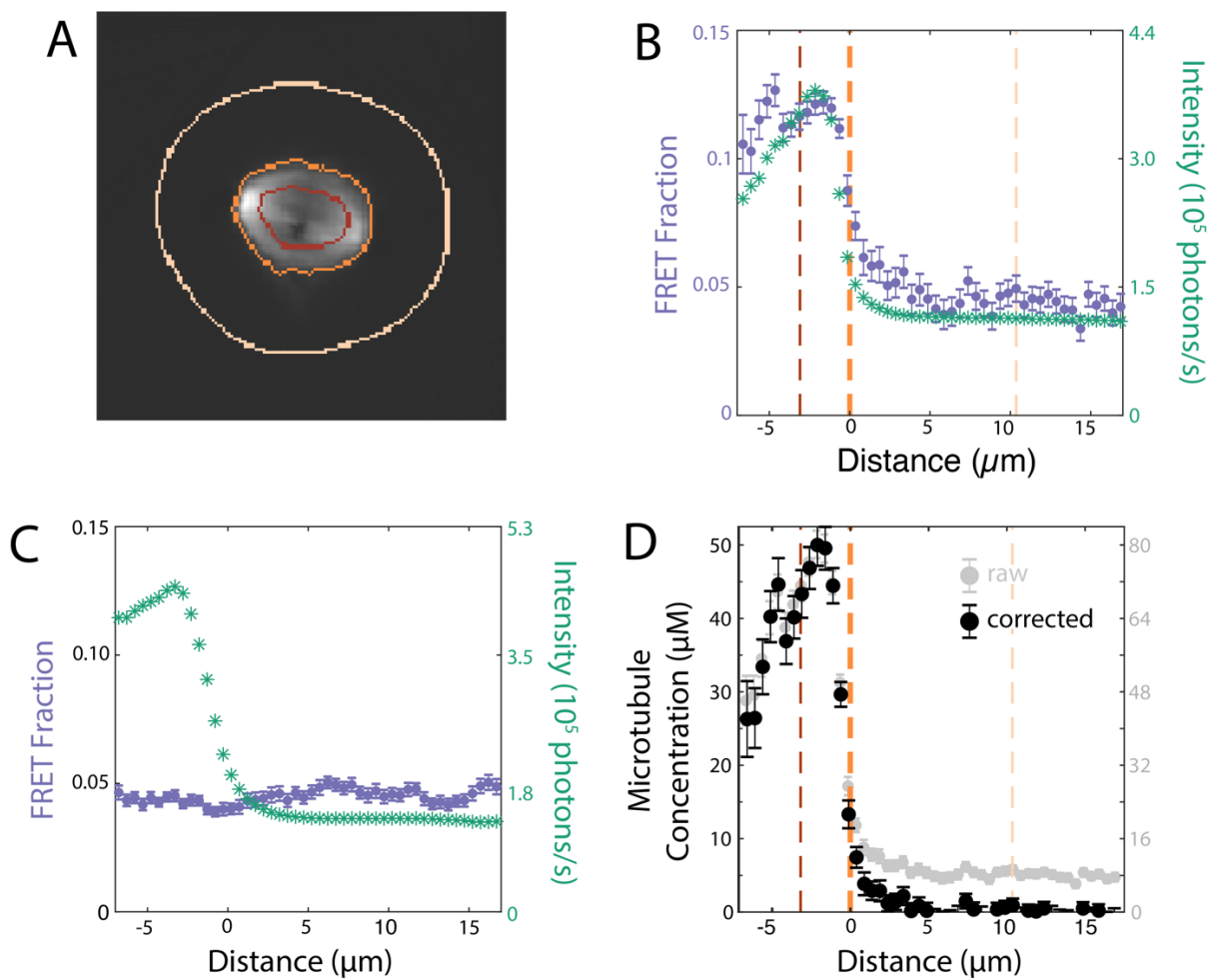

*Figure 3: Polymer concentration versus distance from spindle boundary. (A) Intensity image of spindle. The spindle boundary is shown in orange. Pixels are segmented into groups by their*

*minimum distance from the spindle boundary. Select groupings are shown in light orange and dark orange. (B) FRET fraction (purple circles) and intensity (green stars) versus distance from grouped pixels. Errorbars in FRET fraction are the standard deviation of the posterior distribution. Equidistant pixels are grouped together and the FRET fraction and intensity from each grouping is shown. Pixel groupings corresponding to –3.1 and +10.4 µm are marked by light orange and dark orange dashed lines, respectively. (C) FRET fraction and intensity versus distance from spindles (n=3) formed without acceptor-labeled tubulin. Errorbars in the FRET fraction are the standard deviation of the posterior distribution. The intensity profile is similar to (B) but FRET fraction is no longer increased in the spindle. This level of FRET fraction is used to estimate the bias in measured FRET fraction in (B). (D) Polymer concentration versus distance without subtracting off bias (grey dots) and with subtracting off bias (black dots). Error bars are the standard deviation of the posterior distribution.*

We next investigated if the shape of the spindle interface varied at different locations along the spindle's surface. We divided the spindle into quadrants (Figure 4A). We grouped pixels by distance from the spindle boundary within each quadrant, as previously described. We calculated the microtubule concentration using Eq. 3 and found the microtubule concentration gradient in each quadrant to be similar (Figure 3B). This process was repeated for 11 spindles, revealing similar decay profiles for each quadrant (Figure 3C). Comparing the half-width-at-half-max of the microtubule interface measurements from poleward quadrants to lateral quadrants, we found that the gradient length scale was indistinguishable between these regions (t-test, p-value = 0.70). Taken together, these results argue that the microtubule gradient outside the spindle has little to no dependence on the angle from the long axis of the spindle. Thus, we grouped pixels solely by their distance from the spindle boundary and calculated the microtubule and monomer concentration at each distance from the spindle boundary using Eqs. 3 and 4, respectively. Since all the spindles revealed very similar microtubule gradients, we averaged the microtubule and monomer measurements from each spindle (Figure 4D). The microtubule concentration quickly decays from 58.0µM ± 6.1µM to 0.8µM ± 0.9µM at 5µm where error is estimated as standard error of the mean. The average microtubule concentration at distances larger than 5µm is 0.9µM ± 0.4µM, where error is estimated as the standard deviation of the concentration measurements between 5µm and 15µm. Monomer concentration is depleted in the spindle and levels out approximately where the microtubule and monomer concentration cross. Since we obtained the concentration of tubulin in microtubules, we then sought to construct models to investigate the expected gradient of microtubule concentration for different scenarios of nucleator activity.

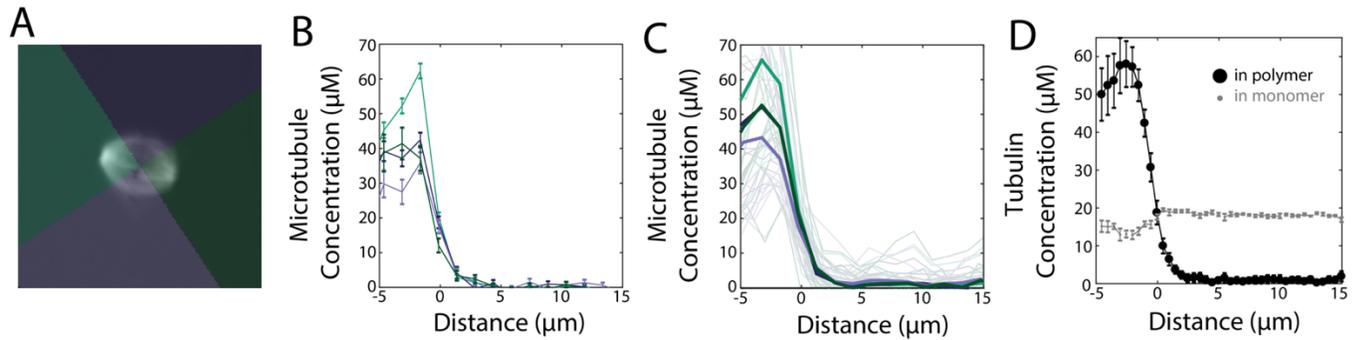

*Figure 4: Polymer gradients decay steeply and uniformly from the spindle boundaries. (A) Intensity image of the spindle shown in figure 3 has been split up into quadrants. Two quadrants correspond to the poles (light green and dark green) and two correspond to the mid body (dark purple and purple). Pixels are segmented into groups by their minimum distance from the spindle boundary in each quadrant. (B) Polymer curves from each quadrant shown in (A). Each quadrant is similar. Errorbars are the standard deviation of the posterior distribution. (C) Polymer curves from each quadrant from 11 spindles. The averaged (opaque curves) and individual (non-opaque) polymer curves show no obvious dependence on quadrant. (D) Microtubule concentration (large black circles) and the monomer concentration (small grey circles) as a function of distance from the spindle boundary are found using the FLIM/FRET technique described above. Results shown are the average of N=11 spindles. Error bars are standard error of the mean.*

## 3 Spindle Interfaces from Reaction Diffusion Dynamics

We next sought to better understand the measured shape of spindle interfaces. To this end, we formulated a model of the spindle interface based on the reaction-diffusion dynamics. In this section, we describe this model starting with a summary of the biochemical pathway of microtubule nucleation.

### 3.1 Chromosome-dependent microtubule nucleation

In metaphase Xenopus spindles, proteins in the nucleation pathway are inhibited from functioning by binding to importin*s*. RanGTP interacts with these complexes, releasing the proteins and triggering microtubule nucleation (Ohba et al. 1999; Zhang, Hughes, and Clarke 1999; Kalab, Pu, and Dasso 1999; Melchior 2001). Ran is a small GTPase, which exists in either a GDP- or a GTP-bound state. RanGDP is converted to RanGTP by the GEF RCC1 at chromosomes, while RanGTP is converted to RanGDP by RanGAP1 in the cytoplasm. This gives rise to a narrow ($\simeq 3\mu$m) RanGTP gradient around the chromosomes (Caudron et al. 2005; Oh, Yu, and Needleman 2016). This gradient has been speculated to give rise to the length scale of the spindle; however, recent experiments have shown that perturbing the length scale of the Ran gradient does not significantly alter the size of the spindle (Oh, Yu, and Needleman 2016). While RanGTP is essential for microtubule nucleation, it is not known whether there is a second level of regulation, namely whether nucleators need to bind to preexisting microtubules in order to nucleate.

Closely following (Oh, Yu, and Needleman 2016), we summarize these findings in a simple model for the dynamics of activated nucleators, denoted $c_u$ and $c_b$ if they are unbound or bound, respectively. In the bulk of the spindle, i.e. far from the RanGTP-enriched region around the chromosomes, the model reads

$$\partial_t c_u = D_c \Delta c_u - (\kappa \rho + r_c) c_u + \kappa_1 c_b \quad (5)$$

$$\partial_t c_b = \kappa \rho c_u - (\kappa_1 + r_c) c_b, \quad (6)$$

where $D_c$ is diffusivity of unbound nucleators, and $r_c$ is their rate of rebinding importin and thus deactivating. Furthermore, $\kappa$ and $\kappa_1$ are the binding and unbinding rate of nucleators to microtubules, respectively. Finally $\rho$ is the density of microtubules. Assuming that the binding-unbinding dynamics of nucleators is fast (see App 7), $\kappa \rho c_u = \kappa_1 c_b$ and the dynamics of the total concentration of nucleators $c = c_u + c_b$ can be obtained from Eqs. 5 and 6 and is given by,

$$\partial_t c = D_c \partial_x^2 \frac{c}{1 + \alpha \rho} - r_c c. \quad (7)$$

Here, $\alpha = \kappa / \kappa_1$ is the binding affinity of nucleator for microtubules.

## 3.2 Microtubule equations of motion

We then developed a description of the growth, shrinkage, and diffusion dynamics of microtubules near the spindle interface. In our model, microtubules are nucleated from an initial size of $\varepsilon$, and grow at velocity $v_g$. They stochastically switch to a depolymerizing state at a rate of $r$. Depolymerizing microtubuless shrink with a velocity $v_d$ until they reach their initial size $\varepsilon$ and disappear. All microtubules diffuse with a size-dependent diffusivity $D/\ell \ln(\ell/d)$ where $\ell$ is their current length and $d = 25$nm is the diameter of a microtubule. Given this model, the polymer number densities of growing microtubules, $\psi_g(x, \ell)$, and shrinking microtubules, $\psi_d(x, \ell)$, obey

$$\partial_t \psi_g(x, \ell) = -v_g \partial_\ell \psi_g + \frac{D}{\ell} \ln\left(\frac{\ell}{d}\right) \partial_x^2 \psi_g - r \psi_g \quad (8)$$

and

$$\partial_t \psi_d(x, \ell) = v_d \partial_\ell \psi_d + \frac{D}{\ell} \ln\left(\frac{\ell}{d}\right) \partial_x^2 \psi_d + r \psi_g. \quad (9)$$

The mass density of polymer is given by

$$\rho = \int_\varepsilon^\infty (\psi_g + \psi_d) \ell d\ell \quad (10)$$

and obeys

$$\dot{\rho} = \varepsilon(v_g\psi_g - v_a\psi_a)|_{\ell=\epsilon} + \int_{\varepsilon}^{\infty}(v_g\psi_g - v_a\psi_a)\,d\ell. \quad (11)$$

In Eq. (11) the first term $\varepsilon(v_g\psi_g - v_a\psi_a)|_{\ell=\epsilon}$ captures the gain and loss of polymer mass by nucleation and disassembly of microtubules, while the second term $\int_{\varepsilon}^{\infty}(v_g\psi_g - v_a\psi_a)\,d\ell$ describes the effect of (de)polymerization dynamics on the microtubule mass. Nucleation enters our equations of motion via the boundary condition

$$\varepsilon v_g \psi_g(x,\ell)|_{\ell=\varepsilon} = m(x), \quad (12)$$

where the locally nucleated microtubule mass $m(x)$ obeys

$$m(x) = c\frac{n_u + \alpha\rho n_b}{1 + \alpha\rho} \quad (13)$$

where $c$ is density of nucleators, $n_b$ is the rate of nucleation by bound nucleators and $n_u$ is the rate of nucleation by unbound nucleators.

## 3.3 Spindle Boundary

To find the steady states of Eqs. (7,8,9), we applied the boundary conditions at the spindle interface and at infinity. Far from the spindle, all nucleators are presumed inactive, and thus

$$c_\infty = 0 \quad (14)$$

This implies that any microtubules far from the spindle originated from spontaneous nucleation - which we take to be zero here - and thus

$$\varepsilon v_g \psi_g|_{x\to\infty,\ell=\varepsilon} = 0 \quad (15)$$

and sets,

$$\psi_g^\infty = \psi_d^\infty = 0 \quad (16)$$

At the other boundary, in the spindle, we set $\rho_s = 1$, which normalizes all densities, such that

$$\psi_g^s = \frac{r^2 v_d}{v_g(\epsilon r + v_g)(v_d + v_g)}\exp\left(-\frac{r}{v_g}(\ell - \epsilon)\right) \quad (17)$$

$$\psi_d^s = \frac{r^2}{(\epsilon r + v_g)(v_d + v_g)}\exp\left(-\frac{r}{v_g}(\ell - \epsilon)\right) \quad (18)$$

Using Eq.12, we solve for the concentration of nucleators in the spindle and find

$$c_s = \frac{\epsilon r^2 v_d(\alpha + 1)}{(v_d + v_g)(\alpha n_b + n_u)(\epsilon r + v_g)}. \quad (19)$$

## 3.4 Microtubule activated nucleation best explains experiments

We next sought to use our model to answer the main question of this paper: are nucleators activated by binding to pre-existing microtubules, or are they active irrespective of their binding state?

We fixed parameters of the model, using measurements from the literature whenever possible. The dynamics of microtubules in the spindle are well characterized. Using the average lifetime of microtubules, we infer the average rate of switching from polymerizing to depolymerizing to be $r = 1/17 \text{s}^{-1}$ (Brugués et al. 2012). In conjunction with the average microtubule length of $6\mu$m, we estimate the microtubule growth velocity to be $v_g = 0.3\mu$m/s. From measurements taken in (Brugués et al. 2012), we set the microtubule depolymerization velocity to be $v_d = 0.6\mu$m/s.

The properties of nucleators are poorly understood and even the proteins responsible for nucleation are still under debate. Here we do not commit to a specific nucleator. For the nucleators diffusivity we choose $D_c = 2\mu\text{m}^2/\text{s}$, which corresponds to the diffusion of $\gamma$TuRC (Lippincott-Schwartz, Snapp, and Kenworthy 2001) in the cytoplasm. This is a reasonable lower bound and places us in a limit where differences between our two models will be least pronounced. For the binding affinity of the nucleator to microtubules, we take the value estimated for the molecule HSET in (Oh, Yu, and Needleman 2016) of $\alpha = 0.024\mu\text{M}^{-1}$.

We set $r_c = 5/4r$, the rate at which the nucleator becomes deactivated, slightly higher than the microtubule catastrophe rate. This value produces the best fit for the polymer gradient shape at the interface. Finally, we set the smallest size of a microtubule $\epsilon = 50$nm, which is about twice the microtubule's diameter, and set this smallest piece to diffuse with a $D_{MT} = \epsilon D \ln(\epsilon/d) = D_c/10$.

Using these parameters, we compared the polymer and monomer gradients shapes at the interface generated by the model to experimental measurements. We measured the microtubule and monomer concentration gradients in 11 spindles, and averaged these results together (Figure 5 A and B, black line). We tested two limiting cases $n_u/n_b = 1$ (Indiscriminate nucleation, green curve) and $n_u/n_b = 0$ (Microtubule activated nucleation, blue curve.) We find that the polymer concentration profiles are better approximated by the microtubule activated nucleation model, see Figure 5 A. From the measured polymer concentrations we furthermore determined the predicted monomer concentration profiles using the fact that the total flux of tubulin vanishes in steady state, i.e. $c_m = -\frac{D}{D_m}\int_\epsilon^{+\infty} \ln\left(\frac{\ell}{d}\right)(\psi_g + \psi_d)d\ell + c_m^\infty$, where $c_m$ and $D_m$ pertain to the concentration and diffusivity of tubulin monomer, respectively. Taking $c_m^\infty$, the monomer concentration at infinity, directly from data and using $D_m = D_{MT}/1.3$, which best fits the data, we find that the experimental monomer concentration profiles as well are best approximated by the microtubule activated nucleation model, see Figure 5 B.

To test the robustness of these findings, we explored a range of parameters around our estimated values and quantified how the half-width-at-half-maximum ($HWHM$, see Figure 5C) of interfaces in experiments would change upon tuning parameters of the model. This is displayed in Fig 5 D to I, in which the colored solid lines display the $HWHM$ for different ratios $n_u/n_b$. The dashed lines mark the parameter estimate used in (A) and (B) and the HWHM estimated from experimental data. Tuning all parameters of the model between half and twice initial estimates we find that robustly microtubule dependent nucleation predicts the sharpest interface. It is noteworthy that the difference between the two models gets less pronounced as the binding affinity of nucleator to microtubules $\alpha$ increases, simply because unbound nucleators become rare for high $\alpha$. However, both models stay distinct until $\alpha$ increases to about an order of magnitude higher than its value estimated in (Oh, Yu, and Needleman 2016). Finally, we tested whether our conclusions were robust to other effects that microtubule binding could have on the behavior of nucleators. In particular, we considered what would happen if microtubule binding changed the lifetime of the nucleator, i.e. bound nucleators might tend to deactivate less than unbound ones. We reformulated our model to incorporate this possibility and did not find a marked difference between the two cases, see App.B. Taken together, our results strongly suggest that nucleators are activated by preexisting microtubules.

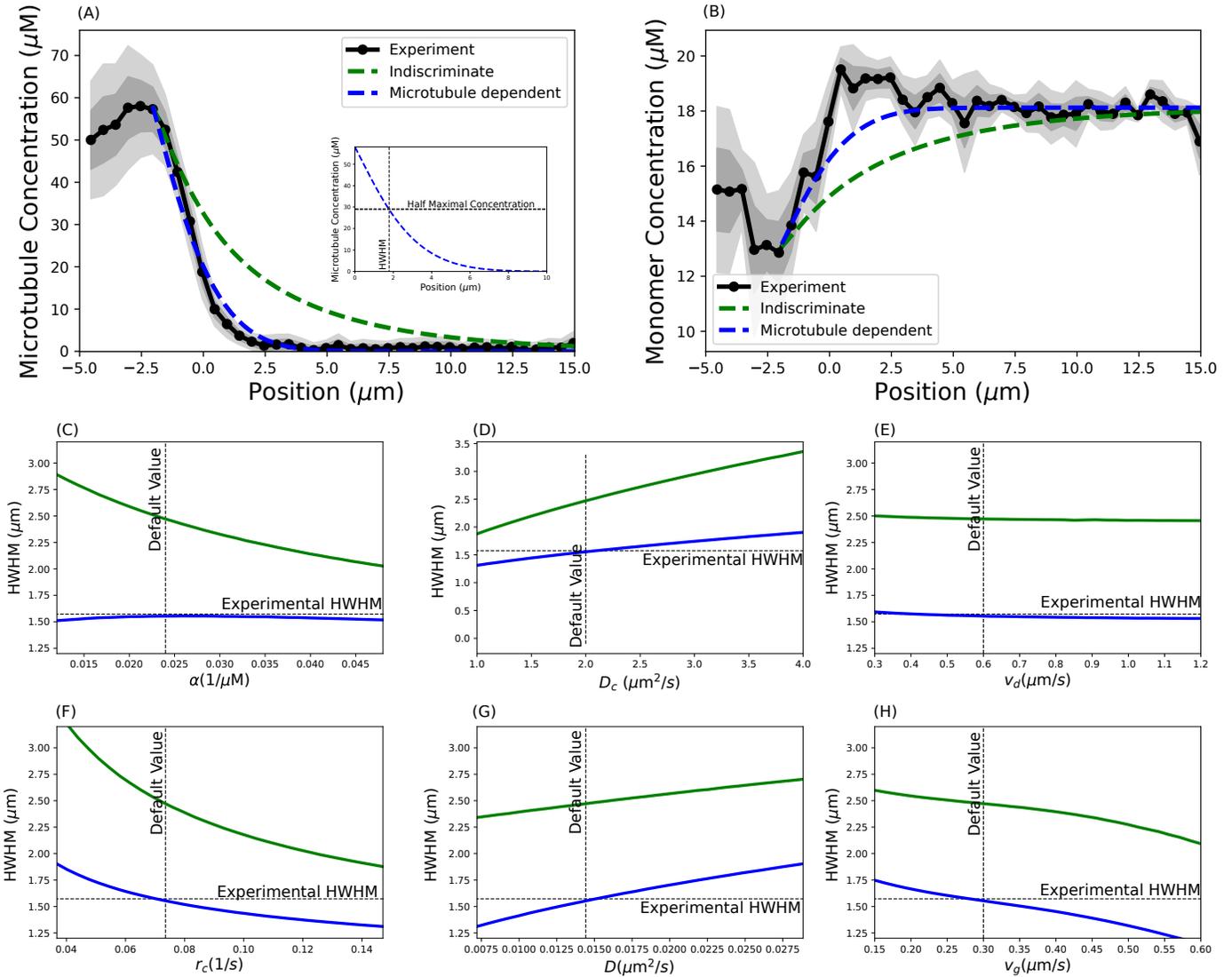

*Figure 5: Comparisons of theory (dashed lines) and experiment (solid black line) for the concentration of tubulin in microtubules (A) and in monomer (B). The blue and green dashed lines show the limiting cases of $n_u = 0$ and $n_u = n_b$, respectively. One and two standard deviations of the data are shown in the dark and light grey regions, respectively. Microtubule activated nucleation (blue) much better describes the data than indiscriminate nucleation (green). We test the robustness of this finding by plotting the half-width-at-half-maximum of the interface HWHW defined in (insert A), for changing parameters $\alpha, D_c, v_g, r_c, D, v_g$ in (C-H), respectively. In the plots, the dotted lines show the parameters used in (A,B) and the HWHM found in experiment.*

## 4 Discussion

Spindle microtubules are nucleated by accessory proteins whose activity is spatially regulated. In this paper, we investigated the spindle interface to learn about the microtubule nucleation mechanism which maintains the spindle as a well-separated structure in the cell. In particular, we investigated whether nucleators binding to preexisting microtubules stimulate their activity.

To do this, we utilized a polymer measurement technique previously described (Kaye et al. 2017) to measure microtubule concentration around the spindle boundary. We measured the microtubule gradient in 11 spindles and found that the microtubule concentration decays sharply at the boundary. The length scale of the decay was indistinguishable between spindles and between poleward and mid-body quadrants of each spindle.

To interpret these results, we formulated a model for diffusing and growing microtubules near the spindle interface. By comparing the predicted and measured microtubule gradients at the spindle interface, we found that the data was consistent with nucleators being activated by binding to preexisting microtubules and inconsistent with nucleators being unaffected by their binding state. We conclude that the rate of microtubule nucleation increases when nucleators bind microtubules. It is noteworthy that the measured microtubule gradient is steeper than the model predicts, and unlike the model, it has both concave and convex parts. This discrepancy may be due to microtubule interactions in the spindle, due to motor proteins and other crosslinkers, which are not included in the model. Previous work has shown that dynein can exert isotropic contractile stresses in metaphase extracts, which could explain the difference between the model and the data (Foster et al. 2015, Foster et al. 2017).

The conclusions of this study are based on measuring and modeling gradients of microtubules and tubulin in and around spindles, and will hold independent of the precise biochemical pathway of nucleation, which is yet to be established. The very sharp gradient of microtubules at the spindle interface argues that the activity of microtubule nucleators is strongly enhanced upon binding microtubules. As microtubule nucleators must first be activated by the Ran pathway (Ohba et al. 1999; Zhang, Hughes, and Clarke 1999; Kalab, Pu, and Dasso 1999), this suggests that the activation of nucleators is a two-step process: inactive nucleators in the cytoplasm are first primed by proximity to chromosomes (by the Ran pathway), but only become fully activated after the primed nucleators bind to microtubules (Clausen et al. 2007; Goshima et al. 2008;  Decker et al. 2017). An important challenge for the future is to establish the molecular basis by which microtubules activate microtubule nucleators.

Different organelles in cells maintain chemically and mechanically distinct micro-environments, even though many of them, like the spindle, are not enclosed by a membrane to separate them from their surroundings. One possibility of maintaining such distinct structures is by providing a scaffold via spatially regulated nucleation, in which the nucleation product feeds back on the activity of the nucleator itself.

Acknowledgements:

We thank Jess Crossno, Emily Davis, and Jan Brugues for thoughtful discussion. B.K. thanks FAS Division of Science, Research Computing Group at Harvard University for access to the Odyssey cluster. B.K. was supported by National Science Foundation GRFP Fellowship DGE1144152. D.N. acknowledges the Kavli Institute for Bionano Science and Technology at Harvard University, United States–Israel Binational Science Foundation Grant BSF 2009271, and National Science Foundation grants PHY-1305254, PHY-0847188, DMR-0820484, and DBI-0959721. Furthermore, this work was partially funded by NSF Grants DMR-1420073 (NYU MRSEC: M.S.), DMS-1463962 (M.S.), DMS-1620331 (M.S.), NIH Grant GM104976 (M.S & D.N.)

# Appendix

## A. Experimental Methods

### Sample Preparation

Samples were measured in a conventional flow cell sealed by candlewax. Bovine tubulin was purified and then labeled with fluorophores as previously described (Mitchison and Kirschner 1984; Hyman et al., 1991, Mitchison lab, 2012). Spindles were assembled in Xenopus laevis egg extracts as previously described (Hannak and Heald, 2006). Tubulin was added to egg extracts by adding donor-labeled tubulin to 0.6 µM and acceptor-labeled tubulin to 1.9 µM. Atto565 was used as the donor fluorophore and Atto647N was used as the acceptor fluorophore.

### Microscopy

Our microscope system was constructed around an inverted microscope (Eclipse Ti, Nikon, Tokyo, Japan) with a commercial scanning system (DCS-120, Becker & Hickl, Berlin, Germany). Fluorophores are excited with a Ti:sapphire pulsed laser (Mai-Tai, Spectra-Physics, Mountain View, CA) at a 1000 nm wavelength, 80 MHz repetition rate (70 fs pulse width), and emitted photons are detected with hybrid detectors (HPM-100-40, Becker & Hickl). The excitation laser was collimated by a telescope assembly to fully utilize the numerical aperture of a water-immersion objective (CFI Apo 40 WI, NA 1.25, Nikon) and avoid power loss at the XY galvanometric mirror scanner. The fluorescence from samples was imaged with a non-descanned detection scheme with a dichroic mirror (705 LP, Semrock) to allow the excitation laser beam to excite the sample while fluorescence passed into the detector path. A short-pass filter was used to further block the excitation laser beam (720 SP, Semrock), followed by an emission filter appropriate for Atto565-labeled tubulin (590/30nm BP, Semrock).

## Acquiring photon arrival-time histograms

We use a Becker and Hickl Simple-Tau 150 FLIM system to collect photon arrival-time histograms. Arrival times are measured relative to an electric pulse created by a photodiode that is triggered by the excitation laser (Becker, 2010). The TAC range was set to 7 x 10-8, with a gain of 5, corresponding to a 14 ns maximum arrival time. The TCSPC system can lose fidelity for photons that arrive just before or after the excitation of the photodiode (Becker, 2010), and thus we set the lower and upper limits to 10.59 and 77.25, respectively, resulting in a 10 ns recording interval. The instrument response was measured using fixed-point illumination of second harmonic generation of a urea crystal. The intensity of the illumination beam was set such that there was an average of 200,000-300,000 photons per second recorded. Data was acquired as a 128x128 pixel image, where a corresponding photon arrival-time histogram was recorded for each pixel.

## Data analysis

Estimating the FRET fraction and Intensity from Photon Arrival Histograms: We use a Bayesian model to build posterior distributions from photon arrival-time histograms (Kaye et al. 2017). The posterior was evaluated at uniformly spaced grid points in parameter space. Point estimates of the FRET fraction were found by taking the maximum of the posterior distribution of the FRET fraction. To reduce the number of free parameters when analyzing photon arrival histograms to find the FRET fraction, we first found the two lifetimes of the donor fluorophore, and then fixed those lifetimes in our Bayesian analysis. The intensity was corrected for inhomogeneous illumination intensity.

Image Registration: To collect a sufficient number of photons for FLIM analysis of FRET, spindles were required to be imaged for 100 seconds. An acquisition for this duration produces blurry images. Thus, we acquired multiple 10-second acquisitions of spindles and aligned the acquisitions as previously described (Brugués et al., 2010). In short, intensity images are thresholded to include the spindle region. The resulting images are then translated so that the center of mass is centered within the image. Each image is then rotated such that the normalized autocorrelation with the previous frame is maximized. After rotation, the images are translated once more such that the normalized autocorrelation is maximized. Translation and rotation were done using the MATLAB (R2017a) function imtranslate (with linear interpolation) and imrotation (with no interpolation), respectively.

Image Segmentation: Registered images were segmented by thresholding the spindle to find the boundary of the spindle. Pixels are then segmented into groups by the shortest distance between the pixel and the spindle boundary. Pixels inside the boundary are considered to have a negative distance and pixels outside the spindle are considered to have a positive distance. The photon arrival-time histograms corresponding to each pixel in a group were added together to create the photon arrival-time histogram corresponding to that distance from the spindle boundary. The intensity of this group is calculated as the mean intensity of the pixels in the group.

Finding $b$ in Eqs. 3 and 4: We solve for $b$, the brightness per µM tubulin, by setting $N_{mon}$ far from the spindle (>10 µm from the spindle boundary) to be 18µM. This calculation assumes that the amount of polymer in this region is negligible, as is consistent with our findings that polymer concentration is indistinguishable from 0 in this region (Figure 3D and 5D). When we do not make this assumption by setting $N_{pol}$ and $N_{mon}$ equal to 18M far from the spindle, we see very similar polymer and monomer curves.

Estimating an upper bound on microtubule concentration outside the spindle: Outside the spindle is defined as at least 5µm from the spindle boundary. We performed Gaussian error propagation on the FRET measurements in figure 3B and 3C to find the mean and standard deviation of the microtubule concentration outside of the spindle. The upper bound was defined as two standard deviations from the mean.

## B. Binding dependent nucleator deactivation

To further probe the robustness of our model we formulated an extension in which we allowed the deactivation of nucleators to depend on their binding state. In this extended model

$$\partial_t c_u = D_c \Delta c_u - (\kappa \rho + r_c) c_u + \kappa_1 c_b \quad (B1)$$

$$\partial_t c_b = \kappa \rho c_u - \kappa_1 c_b - r_c (1 - \beta) c_b, \quad (B2)$$

where $\beta$ is dimensionless and varies from 0 to 1 which describes whether bound nucleators are protected from being deactivated ($\beta = 1$) or turn over like unbound ones ($\beta = 0$). Assuming, like for the model in the main text, that the binding-unbinding dynamics of nucleators is fast (see App C), $\kappa \rho c_u = \kappa_1 c_b$ and the dynamics of the total concentration of nucleators $c = c_u + c_b$ can be obtained from Eqs. B2 and B1 and is given by,

$$\partial_t c = D_c \partial_x^2 \frac{c}{1 + \alpha \rho} - r_c \left(1 - \frac{\beta}{1 + \alpha \rho}\right) c. \quad (B3)$$

In figure Fig (6) we show that regardless of $\beta$ our finding that the interfaces shape is consistent with MT activated nucleation, but not with MT independent nucleation.

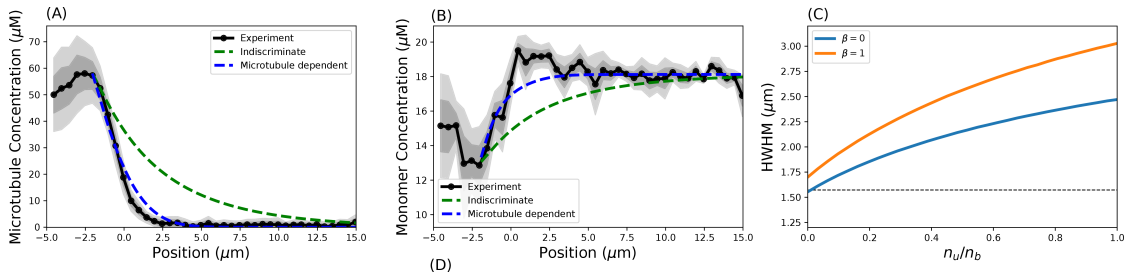

Comparisons of theory (dashed lines) and experiment (solid black line) for the polymer concentration (A) and the monomer concentration (B), for $\beta = 1$. The blue and green dashed lines show the limiting cases of $n_u = 0$ and $n_u = n_b$, respectively. One and two standard deviations of the data are shown in the dark and light grey regions, respectively. The case of

*microtubule activated nucleation (blue) much better describes the data than indiscriminate nucleation (green). Irrespective of β the microtubule activated nuceation produces the sharpest HWHW (C), which corresponds most closely to experiments.*

## C. Fast binding limit of nucleation

To derive the fast time scale limit of our nucleation model we introduce $P = \alpha\rho c_u - c_b$, which obeys, according Eqs. 5 and 6

$$\partial_t P = \alpha\rho D \Delta c_u - (\kappa\rho + \kappa_1 + r_c)P,$$

which upon introducing $\varepsilon = \frac{1}{\kappa\rho + \kappa_1 + r_c}$ can be rewritten as

$$\partial_t \left(e^{-\frac{t}{\varepsilon}} P\right) = e^{-\frac{t}{\varepsilon}} \alpha\rho D \Delta c_u$$

and is solved by

$$P(t) = e^{-\frac{t}{\varepsilon}} P_{-\infty} + \varepsilon \int_{-\infty}^{t} \frac{ds}{\varepsilon} e^{-\frac{(t-s)}{\varepsilon}} \alpha\rho D \Delta c_u \approx 0 + \mathcal{O}(\varepsilon).$$

Thus, in the limit $\kappa\rho + \kappa_1 + r_c \to \infty$

$$\kappa\rho c_u = \kappa_1 c_b.$$

## D. Numerical implementation of the model

We numerically determine the shapes of the interface of the spindle. Our procedure is as follows. Given a spindle density $\rho_1$ we determine a corresponding nucleator density $c_1$ by solving equation 7 with the boundary conditions 19 and 14 using a second order finite difference scheme. Given $c_1$ and $\rho_1$ the spatial distribution of nucleation events $M_1(x)$ is known from Eq.13. We next determine the distribution $\psi_g^{(1)}$, by integrating Eq.8 along the $\ell$ direction, using Eq.12 to set the initial condition at $\ell = \varepsilon$. The integration is performed using scipys odeint integration routine. Using the same technique we also determine $\psi_d^{(1)}$. The steady state solution obeys the fixed point equation

$$F(\rho_1) = \int_{\varepsilon}^{\infty} \ell \left(\psi_g^{(1)} + \psi_d^{(1)}\right) d\ell - \rho_1 = 0. \quad (D6)$$

We find the roots of $F(\rho)$ using Broyden's method.

To compare numerical and experimental data, we align them by defining as the spindle's boundary the first experimental data point for which the polymer density starts decreasing.